\documentclass{elsart}
\usepackage{amsmath,amssymb,graphicx,bm}
\begin{document}

\begin{frontmatter}

\title{
Evolution of Temperature in the Ultracold Strongly-Correlated Plasmas
}

\author{Yurii V.\ Dumin\corauthref{cor}}
\address{
Theoretical Department, IZMIRAN, Russian Academy of Sciences,\\
142190 Troitsk, Moscow region, Russia}
\ead{dumin@yahoo.com}
\corauth[cor]{Corresponding author.}

\begin{abstract}
We present a theoretical interpretation of the recently revealed
features of temperature evolution in the ultracold plasma clouds
released from a magneto-optical trap, namely:
(a)~its independence at the sufficiently large times on
the initial plasma parameters and
(b)~the asymptotics close to~$t^{-1}$ instead of~$t^{-2}$,
expected for a rarefied ideal gas. It is shown that both
these properties can be well explained by the model of
virialization of the charged particle velocities in the regime
of strong electron--ion correlations, while heating due to
inelastic processes (\textit{e.g.} three-body recombination)
should be of secondary importance. These conclusions are confirmed
also by the results of \textit{ab initio} computer simulations.
\end{abstract}

\end{frontmatter}

\section{Introduction}

A new interesting branch of plasma physics, emerged in the last
decade, is studying the clouds of rarefied ultracold plasmas
produced by the laser capture and cooling of gases in
the magneto--optical traps (MOT) and their subsequent ionization
(for general review see, for example,~%
\cite{bib:PhysTod_Gould,bib:PhysTod_Berg,bib:PhysRep_Killian}
and references therein).
These are the classical (non-quantum) gaseous systems with
characteristic temperatures from a fraction to a few Kelvin, in which
the Coulomb's coupling parameter~$ \Gamma = e^2 n^{1/3} / k_B T $
can reach considerable values. For example, the measured values of
this parameter for the ions~$ {\Gamma}_i $ are
about~$ 2{\div}3 $~\cite{bib:PRL_Killian}; the estimates
of~$ {\Gamma}_e $ for the electrons are less accurate and model
dependent, but they also results in the values comparable to
unity.

The possibility of existence of such metastable plasmas was
theoretically predicted a quite long time ago (\textit{e.g.}
article~\cite{bib:JETPL_TY} and references therein), but
they were created experimentally only after sufficient development
of the technology for laser cooling of atoms. Besides,
it was proposed to produce similar plasma states by the artificial
release of the ionized gas clouds from
spacecraft~\cite{bib:JLTP_Dumin,bib:APSS_Dumin}, but the possibility
of diagnostics in cosmic space remains too limited by now.

One of the most interesting recent results in the experimental study
of ultracold plasmas is the behavior of temperature in the clouds
released from a trap and expanding freely in space.
It was unexpectedly found that, firstly, the law of decrease of
the electron temperature became universal at large times,
\textit{i.e.} did not depend on the initial conditions.
(For example, when the initial temperatures~$ T_e $ varied by
30~times, their difference after a few microseconds was only
$ 2{\div}3 $~times and decreased further in the course of
the subsequent evolution~\cite{bib:PRL_Roberts}.) Secondly,
which is even more interesting, the asymptotics was measured to be
$ T_e \propto t^{-(1.2 \pm 0.1)} \approx t^{-1} $%
~\cite{bib:PRL_Fletcher_2007} instead of~$ t^{-2} $, which
would be expected for an ideal rarefied gas without internal degrees
of freedom ($ \gamma \! = \! 5/3 $) at the inertial stage of its
expansion (\textit{i.e.} when the plasma cloud expands with
a constant velocity, so that its characteristic size increases
linearly with time, $ R \propto t $).

The most straightforward way to interpret the substantially slower
decrease in the electron temperature is to take into account
a heat release by the recombination of charged particles.
In the particular case of atomic ions, the most efficient channel
should be three-body recombination $ A^+ + e + e \to A + e $,
when one electron is captured by the ion, while the second electron
carries away the excessive energy. Unfortunately, the recent attempts
of numerical simulation of the observed law of temperature evolution
with the three-body recombination did not lead to the satisfactory
results; see, for example, Fig.~3a~\cite{bib:PRL_Fletcher_2007}).%
\footnote{
Let us mention that the method of plotting the temperature curves
in Fig.~3a~\cite{bib:PRL_Fletcher_2007} is slightly confusing.
According to the physical sense of the problem, the various laws of
evolution should be compared at the same initial temperature;
while in the above-mentioned figure they were presented at various
initial temperatures. As a result, at first sight, the disagreement
between the curves appears at small rather than at large times.}

The aim of our paper is to show that all the experimentally observed
features of temperature evolution (namely, both the establishment of
the universal asymptotics, independent of the initial conditions,
and its particular form, close to~$ t^{-1} $) can be naturally
explained by the model of virialization of the charged particle
velocities, \textit{i.e.} actually due to changing the equation of
state of the ultracold plasma under the presence of strong
electron--ion correlations. Thereby, in the first approximation,
it is not necessary to take into account any inelastic processes,
such as heat release by the three-body recombination.

\section{Theoretical Model}

Our analysis will consist of the three basic steps. First of all,
let us consider a sufficiently small (but macroscopic) element
of the expanding plasma where thermodynamic equilibrium is supposed
to be established. Then, we can describe it by the multi-particle
distribution function of the following general form:
\begin{align}
& f( \textbf{r}_{e 1}, \dots , \textbf{r}_{e N_e},
\textbf{v}_{e 1}, \dots , \textbf{v}_{e N_e} ) =
A_f \exp \! \Big\{
  \!\! - \! \frac{1}{k_B T_e}
\nonumber \\
& \quad \times \Big[
  \, \sum_{n=1}^{N_e} \, \frac{ m_e \textbf{v}_{en}^2 }{ 2 }
  + \, U( \textbf{r}_{e 1}, \dots , \textbf{r}_{e N_e},
  \textbf{r}_{i 1}, \dots , \textbf{r}_{i N_i} )
\Big] \! \Big\} \, ,
\end{align}
where $ {\bf r}_{en} $ and $ {\bf v}_{en} $ are the coordinates and
velocities of electrons, $ {\bf r}_{in} $ are the coordinates of ions,
and $ A_f $ is the normalization factor.
(Kinetic energy of the ions is ignored here, since it is usually much
less than the kinetic energy of electrons in the particular experimental
setups. In principle, the same consideration can be conducted with the
kinetic energy of ions if the thermodynamic equilibrium is established
between all kinds of the charged particles.)

Despite a very complex form of the potential energy~$U$ in the regime
of strong Coulomb's interaction, the average value of some quantity~$F$
depending only on the velocities~$ {\bf v}_{en} $ can be calculated
quite easily:
\begin{equation}
\langle F(\textbf{v}_e) \rangle = \frac{\displaystyle %
\int \! F(\textbf{v}_e) \exp \Big\{ \!\! - \! \frac{1}{k_B T_e}
  \Big[ \, \sum_{n=1}^{N_e} \, \frac{ m_e \textbf{v}_{en}^2 }{ 2 }
  \Big] \Big\} \, d \textbf{v}_e }{\displaystyle %
\int \! \exp \Big\{ \!\! - \! \frac{1}{k_B T_e} \Big[ \,
  \sum_{n=1}^{N_e} \, \frac{ m_e \textbf{v}_{en}^2 }{ 2 }
  \Big] \Big\} \, d \textbf{v}_e } \; ,
\end{equation}
because the integrals~$ \int \exp \{ - U (\textbf{r}_e ,
\textbf{r}_i) / k_B T_e \} \, d \textbf{r}_e $
in the numerator and denominator exactly cancel each other.
(Here, $ \textbf{v}_e $, $ \textbf{r}_e $, and $ \textbf{r}_i $
designate the entire sets of velocities and coordinates of
the electrons and ions, respectively.)

Therefore, the average kinetic energy per one particle will be
\begin{equation}
\langle k \rangle = (3/2) \, k_B T_e \, .
\end{equation}
This formula looks exactly as for the ideal gas, but it is actually
applicable for a plasma with arbitrary strong Coulomb's interaction~$U$
between the particles.

Next, at the second step of our consideration, the average kinetic
energy~$ \langle k \rangle $ can be related to the average potential
energy~$ \langle u \rangle $ by the virial theorem for the Coulomb's
field~\cite{bib:Landau_1976}:%
\footnote{
Since the virial theorem directly relates the energies averaged over
time, we need to assume also that the system is ergodic, \textit{i.e.}
the quantities averaged over time are equal to the ones averaged
over ensemble.}
\begin{equation}
\langle k \rangle = (1/2) \: | \langle u \rangle | \, ,
\end{equation}
which is also valid at the arbitrary strength of interparticle
interaction. Strictly speaking, the virial theorem is applicable only
to the system of particles experiencing a finite
(\textit{i.e.} restricted in space) motion. Nevertheless, we can expect
that this theorem will be sufficiently accurate also for a freely
expanding plasma cloud if the characteristic time of variation in
its macroscopic parameters ($ \gtrsim 10^{-5} $~s for the particular
experimental setup~\cite{bib:PRL_Fletcher_2007,bib:PRL_Fletcher_2006})
is considerably greater than the microscopic periods of motion of
the electrons ($ 10^{-9} \div 10^{-7}$~s).

Finally, at the last step of our consideration, the average potential
energy can be evidently expressed through the characteristic distance
between the particles:
\begin{equation}
\langle u \rangle \sim \, e^2 / \langle r \rangle \sim \,
e^2 n^{1/3} \, .
\end{equation}

Therefore, combining all the above formulas, we
get~$ T_e \propto n^{1/3} $. In particular, if the cloud expands
inertially (\textit{i.e.} linearly in time) and, consequently,
its concentration changes as~$ t^{-3} $, we get:
\begin{equation}
T_e \propto t^{-1} \, .
\end{equation}

In summary, we conclude that the presented model of virialization
of the charged particle velocities well explains both
experimental features of thermal evolution of the plasma, namely:
(a)~the system ``forget'' in the course of time about its
initial temperature, \textit{i.e.} the plasma clouds with
various initial temperatures evolve by the same way; and
(b)~the particular functional dependence is close to~$ t^{-1} $
instead of the intuitively expected~$ t^{-2} $.
The inelastic processes, such as three-body recombination,
are not of importance in this model.

\section{Numerical Simulations}

To verify the above theoretical estimates, we performed
\textit{ab initio} simulation of the plasma dynamics, based on
the numerical solution of the equations of classical mechanics
for the multi-particle system. Our approach differed from
the earlier works in the following aspects.

First of all, the authors of most of the previous simulations
of ultracold plasmas tried to include into consideration as many
particles as possible. Unfortunately, to reduce the computational
cost, they had to use some simplifications of the equations of
motion of the light particles (electrons), such as the
particles-in-cell (PIC) method~\cite{bib:PhysPlasm_Rob} or
Vlasov approximation for the electrons~\cite{bib:PRL_Pohl}.
Both these approaches are based on the introduction of average
electric fields and, therefore, completely ignore strong
individual electron--ion interactions (large-angular scattering),
which are just responsible for the virialization of velocities.
As distinct from these approximations, we did not intend
to simulate as many particles as possible but tried to integrate
the equations of motion of the electrons in the real electric
microfields as accurately as possible, without using any extra
simplifications. (The ions were assumed to be very heavy and moved
by a purely inertial law.)

Yet another well-known problem in modeling of the expanding
plasmas is a considerable change of the spatial scale of the
system (and, therefore, the amplitude of Coulomb's forces) during
its evolution. As a result, it is quite difficult to choose
the method of numerical integration ensuring a reasonable accuracy
in the entire time interval. We resolved this problem by
introducing a ``scalable'' coordinate frame, expanding in space
with the average velocity of plasma outflow. In other words,
the initial equations of the electron motion
\begin{equation}
d^2 {\rm \bf r}_{ek} / dt^2 = \, {\rm \bf F}_{ek} \, , \quad
k = 1, \dots , N_e
\label{eq:motion_diment}
\end{equation}
(where $ {\rm \bf F}_{ek} $ is the total Coulomb's force
acting on $k$'th electron from all other electrons and ions)
after the introduction of dimensionless variables
$ r^* = r / \, \tilde{l} \, , \; t^* = t / \tau $
and transformation to the coordinate frame expanding with
plasma, $ \tilde{l} = \, \tilde{l_0} \, ( 1 + \, u_0^* \, t^* ) $,
can be reduced to
\begin{equation}
{\ddot{\rm \bf r}}_{ek}^*
+ \, 2 \, u_0^* \, ( 1 + \, u^*_0 \, t^* )^{-1} \,
  {\dot{\rm \bf r}}_{ek}^* = \,
( 1 + \, u^*_0 \, t^* )^{-3} \, {\rm \bf F}^*_{ek} \, .
\label{eq:motion_dimless}
\end{equation}
Here, $ \tau = {\left( m / Z e^2 \right)}^{\! 1/2} \:
{\tilde{l_0}}^{\, 3/2} $~is the characteristic plasma time
(on the order of the inverse Langmuir frequency),
$ \tilde{l} $~is the characteristic distance between the particles
($ \tilde{l_0} $~is its value at the initial instant of time);
$ u_0^* $ is the dimensionless velocity of the inertial expansion
of the plasma cloud, determined by the relation
$ u_0^* = u_0 \, \tau / L_0 $, where
$ L_0 $~is the size of the computational cell used in our simulations,
and $ u_0 $~is the velocity of its boundary;
$ {\rm \bf F}^*_{ek} $~are the dimensionless Coulomb's forces,
and the dot denotes derivative with respect to the dimensionless
time~$ t^* $.

Therefore, as follows from equations ~(\ref{eq:motion_dimless}),
the effect of inertial plasma outflow in the expanding coordinate
system looks like an effective dissipative force, proportional to
the electron velocities. Consequently, the temperature of
the electron gas is determined by the competition between two effects:
(a)~acceleration and heating of the electrons due to Coulomb's
interactions with ions and
(b)~their deceleration and cooling by the above-mentioned dissipative
forces.

It is important to emphasize also that transformation from the
equations~(\ref{eq:motion_diment}) to~(\ref{eq:motion_dimless})
enabled us to perform a numerical integration in the fixed region
of dimensionless coordinates and, therefore, to avoid the problem
of quick losing the computational accuracy when the spatial scale
of the system changes very much.

\begin{figure}
\includegraphics[width=8cm]{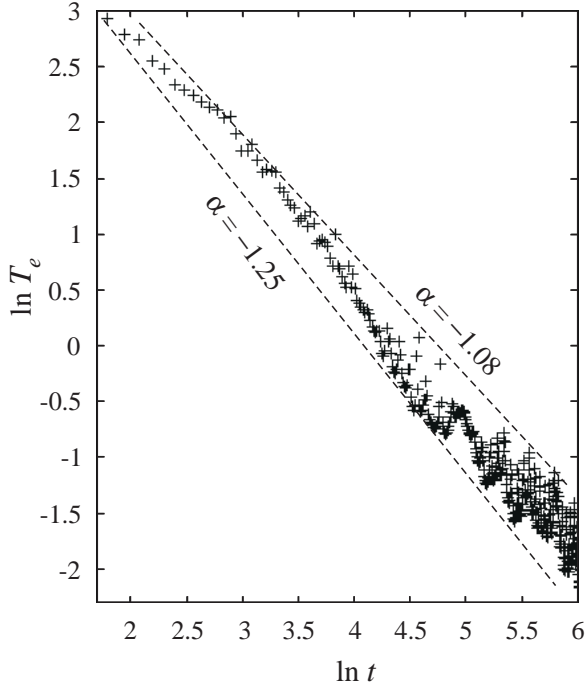}
\caption{\label{fig:Simulations}
Simulated electron temperature as function of time in
the logarithmic scale. The inclined dashed lines show
the power-like dependences~$ T_e \propto t^{\alpha} $; the values
of~$ \alpha $ being presented near the respective lines.
}
\end{figure}

The results of our simulations are presented by crosses in
Fig.~\ref{fig:Simulations}. If we do not take into account
the earliest time interval, when the relaxation processes occur,
the computational points in logarithmic coordinates are located
almost along a straight line, corresponding to the power law
of evolution of the electron temperature: $ T_e \sim t^{\alpha} $.
The corresponding exponent was estimated to be within the limits
$ \alpha = - (1.08 \div 1.25) $. This is quite close to the value
$ \alpha = - 1 $, following from the simple virial estimate,
and is in excellent agreement with the experimental values
$ \alpha = - (1.1 \div 1.3) $~\cite{bib:PRL_Fletcher_2007}.

\section{Conclusions}

Analysis of the available data on the evolution of the electron
temperature in the freely expanding ultracold plasma clouds shows
that a heat release due to three-body recombination cannot explain
quantitatively the experimental results. On the other hand,
taking into account the strong electron--ion correlations and the
resulting virialization of the charged particle velocities
(\textit{i.e.} modification of the equation of state of the plasma)
leads to the perfect agreement with the experimental data.


\begin{thebibliography}{10}
\expandafter\ifx\csname url\endcsname\relax
  \def\url#1{\texttt{#1}}\fi
\expandafter\ifx\csname urlprefix\endcsname\relax\def\urlprefix{URL }\fi

\bibitem{bib:PhysTod_Gould}
P.~Gould, E.~Eyler, Phys. World 14~(3) (2001) 19.

\bibitem{bib:PhysTod_Berg}
S.~Bergeson, T.~Killian, Phys. World 16~(2) (2003) 37.

\bibitem{bib:PhysRep_Killian}
T.~Killian, T.~Pattard, T.~Pohl, J.~Rost, Phys. Rep. 449 (2007) 77.

\bibitem{bib:PRL_Killian}
C.~Simien, Y.~Chen, P.~Gupta, S.~Laha, Y.~Martinez, P.~Mickelson, S.~Nagel,
  T.~Killian, Phys. Rev. Lett. 92 (2004) 143001.

\bibitem{bib:JETPL_TY}
A.~Tkachev, S.~Yakovlenko, JETP Letters 73 (2001) 66.

\bibitem{bib:JLTP_Dumin}
Y.~Dumin, J. Low Temp. Phys. 119 (2000) 377.

\bibitem{bib:APSS_Dumin}
Y.~Dumin, Astrophys. Space Sci. 277 (2001) 139.

\bibitem{bib:PRL_Roberts}
J.~Roberts, C.~Fertig, M.~Lim, S.~Rolston, Phys. Rev. Lett. 92 (2004) 253003.

\bibitem{bib:PRL_Fletcher_2007}
R.~Fletcher, X.~Zhang, S.~Rolston, Phys. Rev. Lett. 99 (2007) 145001.

\bibitem{bib:Landau_1976}
L.~Landau, E.~Lifshitz, Mechanics, Pergamon, Oxford, 1976.

\bibitem{bib:PRL_Fletcher_2006}
R.~Fletcher, X.~Zhang, S.~Rolston, Phys. Rev. Lett. 96 (2006) 105003.

\bibitem{bib:PhysPlasm_Rob}
F.~Robicheaux, J.~Hanson, Phys. Plasm. 10 (2003) 2217.

\bibitem{bib:PRL_Pohl}
T.~Pohl, T.~Pattard, J.~Rost, Phys. Rev. Lett. 92 (2004) 155003.

\end{thebibliography}

\end{document}